\documentstyle[12pt,fleqn]{article}

\def\eps{\varepsilon}

\def\beeq{\begin{equation}}
\def\eneq{\end{equation}}
\def\beeqa{\begin{eqnarray}}
\def\eneqa{\end{eqnarray}}

\def\eps{\varepsilon}

\addtocounter{section}{-1}
\begin{document}

\begin{center}

{\large {\bf Elastic anomaly of heavy fermion systems\\
in a crystalline field}\\
}
{Running head: {\sl Elastic anomaly of heavy fermion systems}}\\

\vspace{1cm}

{\rm Kikuo Harigaya\footnote[2]{Permanent address:
Fundamental Physics Section, Electrotechnical Laboratory,
Umezono 1-1-4, Tsukuba, Ibaraki 305, Japan;
E-mail address: harigaya@etl.go.jp.}
and G. A. Gehring
}\\

\vspace{1cm}

{\sl Department of Physics, University of Sheffield,\\
Sheffield S3 7RH, United Kingdom}\\

\vspace{1cm}

(Received~~~~~~~~~~~~~~~~~~~~~~~~~~~~~~~~~~~)
\end{center}

\vspace{1cm}

\noindent
{\bf Abstract}\\
An elastic anomaly, observed in the heavy Fermi liquid state of Ce
alloys (for example, CeCu$_6$ and CeTe), is analyzed by using the
infinite-$U$ Anderson lattice model.  Four atomic energy levels
are assumed for f-electrons.  Two of them are mutually degenerate.
A small crystalline splitting $2\Delta$ is assumed between two
energy levels.  Fourfold degenerate conduction bands are also
considered.  We solve the model using the mean field approximation
to slave bosons and the renormalized elastic constant is calculated.
The temperature dependence of the constant shows the downward dip.
The result is compared with the experimental data of CeTe.

{}~

\noindent

\pagebreak


\section{Introduction}

Elastic constants of heavy fermion compounds show anomalies
interpreted as the crystalline field effect.  For example,
the elastic constant $c_{33}$ of CeCu$_6$ [1] has a dip at
about 10K.  This might be due to the splitting larger than
the Kondo temperature 4K.  The constant $(c_{11}-c_{12})/2$
of CeTe [2] shows the apparent dip at about 15K.  There is the
crystalline field splitting 30K between 4f-orbitals.  This is
the origin of the dip.

In this paper, we present a microscopic calculation of the elastic
anomaly in order to look at effects of the degeneracy structure and
crystalline field.  We discuss by the mean field theory of the infinite-$U$
Anderson lattice model.  The temperature can be varied from $T=0$ to
$T$ much higher than the Kondo temperature, as discussed previously [3,4].
We assume an empirical relation between an elastic constant and
the susceptibility with respect to the crystalline field [5].
We show the temperature dependence of the elastic constant and
compare with the downward dip observed in the constant
$(c_{11}-c_{12})/2$ of CeTe [2].  We note that the details
of the underlying theory have been published in [6] and this
paper contains the comparison with the experiment.

\section{Model}

We consider the infinite-$U$ Anderson lattice model in the
slave boson method.  The model has the following form:
\beeqa
H &=& \sum_{i}
[ ( E_{\rm f} - \Delta ) \sum_{l=1,2} f_{i,l}^\dagger f_{i,l}
+ ( E_{\rm f} + \Delta ) \sum_{l=3,4} f_{i,l}^\dagger f_{i,l} ] \\ \nonumber
&+& \sum_{{\vec k},l=1-4} \eps_{\vec k} c_{{\vec k},l}^\dagger
c_{{\vec k},l} \\ \nonumber
&+& V \sum_{i,l=1-4} ( f_{i,l}^\dagger c_{i,l} b_i
+ b_i^\dagger c_{i,l}^\dagger f_{i,l} ) \\ \nonumber
&+& \sum_i \lambda_i ( \sum_{l=1-4} f_{i,l}^\dagger f_{i,l}
+ b_i^\dagger b_i - 1),
\eneqa
where $f_{i,l}$ is an f-electron operator
of the $l$-th orbital at the $i$-th site, $c_{{\vec k},l}$ is
a conduction electron operator with the wave number
${\vec k}$, and $b_i$ is a slave boson operator.
The atomic energy of the first and second orbitals of f-electrons is
$E_{\rm f} - \Delta$, and that of the third and fourth orbitals
is $E_{\rm f} + \Delta$.  The magnitude of the crystalline field splitting
is $2 \Delta$.  We use the square density of states, $\rho \equiv 1/ND$,
for the conduction band.  This extends over the
energy region, $-D < \eps_{\vec k} < (N-1) D$ ($N = 4$),
meaning that the combination $N \rho V^2$ which appears
in the $1/N$ expansion is independent of $N$,
and that the mean field theory becomes exact as $N \rightarrow \infty$.
This model is treated within the mean field approximation.

\section{Elastic anomaly in low temperatures}

An elastic constant $c$ is related with the linear susceptibility
with respect to $\Delta$, as shown by the formula [5],
$c = c_0 / (1 + g \chi_\Delta)$,
where $c_0$ is the elastic constant of the system where there is
not lattice-electron interactions,
and $g = c_0 \eta^2$ where $\eta$ is the coupling constant
between $\Delta$ and the strain field.  As the value of $g$ is unknown,
we treat it as a fitting parameter (we note that $g \geq 0$).  The quantity
$\chi_\Delta$ is calculated as the second order derivative of
the mean field free energy:
$\chi_\Delta = - \partial^2 F / \partial \Delta^2$.

We show the temperature dependence of $c/c_0$ for $g = 2.3$K in Fig. 1.
The elastic constant decreases from much higher to lower temperatures
than the Kondo temperature ($\simeq$ 70K).
There is a downward dip around 15-20 K, which is the
result of the combination of the fourfold orbital degeneracy and
the crystalline field splitting.  A larger value of $\Delta$ gives rise
to a larger dip.  The constant $(c_{11}-c_{12})/2$ of CeTe [2]
shows the apparent dip at about 15K.  The experimental data are plotted again
by dots in Fig. 1.  The high temperature behavior, where $1 - c/c_0$
is almost inversely proportional to $T$, shows remarkable agreement.
The structure around the dip is qualitatively explained
by the present calculation.

\section{Discussion}

The downward dip of $c$ exists even if $\Delta = 0$.
The finite $\Delta$ results in the shift of the dip to lower temperatures.
The temperature at the dip would be finally determined by the combination
of the degeneracy of the 4f orbitals and the splitting width.

One of the interesting future problems would be the magnetic field
effects.  Further removal of the degeneracy will give rise to
more structure in the elastic constants.


\begin{flushleft}
{\bf References}
\end{flushleft}

\noindent
$[1]$ T. Goto, T. Suzuki, Y. Ohe, T. Fujimura, S. Sakutsume,
Y. Onuki, and T. Komatsubara, J. Phys. Soc. Jpn. 57 (1988) 2612.\\
$[2]$ H. Matsui, T. Goto, A. Tamaki, T. Fujimura, T. Suzuki, and
T. Kasuya, J. Magn. Magn. Mater. 76/77 (1988) 321.\\
$[3]$ S. M. M. Evans, T. Chung, and G. A. Gehring,
J. Phys.: Condens. Matter 1 (1989) 10473.\\
$[4]$ K. Harigaya, Master Thesis, University of Tokyo (1988).\\
$[5]$ P. Levy, J. Phys. C: Solid State Phys. 6 (1973) 3545.\\
$[6]$ K. Harigaya and G. A. Gehring, J. Phys.: Condens. Matter
5 (1993) 5277.\\


\begin{flushleft}
{\bf Figure Captions}
\end{flushleft}

\noindent
Fig. 1.  Temperature dependences of the elastic constant $c/c_0$
for $g = 2.3$K, compared with the experimental data [2] of CeTe
shown by the dots.  Parameters are $D = 5 \times 10^4$K,
$V = 7500$K, $E_{\rm f} = - 10^4$K, and the electron number
$n_{\rm el} = 1.9$.
The dashed, thin, and heavy lines are for $\Delta = 0$, 15,
and 30K, respectively.

\end{document}